\documentclass[12pt]{article}
\usepackage{graphicx}
\begin{document}

\title{Manifestation of the JLab proton polarization data on the behaviour of strange
       nucleon form factors}

\author{Stanislav Dubni\v cka\\
        Inst. of Physics, Slovak Academy of Sciences, Bratislava, Slovak Republic\\
\medskip
        Anna Zuzana Dubni\v ckov\'a\\
        Dept. of Theoretical Physics, Comenius Univ., Bratislava, Slovak Republic}
\maketitle
\begin{abstract}
    Special eight-resonance unitary and analytic model of nucleon
    electromagnetic structure is used to analyze, first the
    classical proton form factor data obtained by the Rosenbluth
    technique and then also the contradicting JLab proton
    polarization data on the ratio $\mu_p G_{Ep}(Q^2)/G_{Mp}(Q^2)$
    with the aim to investigate a manifestation of the latter on
    the strange nucleon form factors behaviour.

\end{abstract}

\bigskip

    PACS numbers: 12.40.Vv, 13.40.Fn, 14.20.Dh

\bigskip

   Owing to the quark structure of proton and neutron one doesn't
know explicit form of the nucleon matrix element of the
electromagnetic (EM) current $J^{EM}_\mu =2/3\bar u\gamma_\mu u -
1/3\bar d\gamma_\mu d -1/3\bar s\gamma_\mu s$ and as a result two
independent scalar functions, called nucleon EM form factors
(FF's), of the squared four-momentum $t = -Q^2$, transferred by
the exchanged virtual photon are introduced for proton and
neutron, respectively. Very natural is an introduction of Dirac
$F_{1N}(t)$ and Pauli $F_{2N}(t)$ FF's
\begin{equation}
<N\mid J^{EM}_\mu \mid N> = \bar u(p')\{\gamma_\mu
F_{1N}(t)+i\frac{\sigma_{\mu\nu}q_\nu}{2m_N}F_{2N}(t)\}u(p).
\end{equation}
The most suitable in extracting of experimental information are
Sachs electric $G_{EN}(t)$ and magnetic $G_{MN}(t)$ FF's
\begin{equation}
G_{EN}(t)=F_{1N}(t)+\frac{t}{4m_N^2}F_{2N}(t),\quad
G_{MN}(t)=F_{1N}(t)+F_{2N}(t)
\end{equation}
giving in the Breit frame the charge and magnetization
distributions within the nucleon, respectively. However, for a
construction of various models of nucleon EM structure the
isoscalar and isovector Dirac and Pauli FF's, to be defined by the
matrix elements
\begin{equation}
   \langle p^{'}|J^{I=0}_{\mu}|p\rangle=\bar{u}(p^{'})
\left[\gamma_{\mu} F^{I=0}_{1}(t)+
i{{\sigma_{\mu\nu}q^{\nu}\over{2m_N}}} F^{I=0}_{2}(t)\right]u(p)
\label{FF0}
\end{equation}
of the isoscalar EM current $J^{I=0}_{\mu}$={1/6}
$(\bar{u}\gamma_{\mu}u+\bar{d}\gamma_{\mu}d)$-{1/3}$\bar{s}
\gamma_{\mu}s$ and

\begin{equation}
   \langle p^{'}|J^{I=1}_{\mu}|p\rangle=\bar{u}(p')
\left[\gamma_{\mu} F^{I=1}_{1}(t)+
i{{\sigma_{\mu\nu}q^{\nu}\over{2m_N}}} F^{I=1}_{2}(t)\right]u(p)
\label{FF1}
\end{equation}
of the isovector EM current $J^{I=1}_{\mu}$={1/2}$(\bar{u}
\gamma_{\mu}u-\bar{d}\gamma_{\mu}d)$, are the most appropriate.

   One of the important tasks of modern hadron physics is to
clarify the role of hidden flavours in the structure of the
nucleon. The contribution of the strange quarks is of special
interest as their mass is within the range of the mass scale of
QCD ($m_s\approx \Lambda_{QCD}$), so the dynamic creation of sea
strange quark-antiquark pairs could still be dominating in
comparison with heavier $c, b$ and $t$ quark-antiquarks pairs
creation.

   The momentum dependence of the nucleon matrix element of the strange-quark
vector current $J^{s}_{\mu}$=$\bar{s}\gamma_{\mu} s$ is contained
in the Dirac $F^{s}_{1} (t)$ and Pauli $F^{s}_{2} (t)$ strange
nucleon FF's

\begin{equation}
   \langle p^{'}|\bar{s}\gamma_{\mu} s|p\rangle=\bar{u}(p^{'})
\left[\gamma_{\mu} F^{s}_{1}(t)+
i{{\sigma_{\mu\nu}q^\nu}\over{2m_N}} F^{s}_{2}(t)\right]u(p)
\label{FFS}
\end{equation}
or in the strange electric $G_E^s(t)$ and strange magnetic
$G_M^s(t)$ FF's

\begin{equation}
G_{E}^s(t)=F_{1}^s(t)+\frac{t}{4m_N^2}F_{2}^s(t),\quad
G_{M}^s(t)=F_{1}^s(t)+F_{2}^s(t),
\end{equation}
which, as a consequence of the isospin zero value of the strange
quark, contribute only to the behaviour of the isoscalar nucleon
FF's and never to isovector ones.

   Recent measurements \cite{jon}-\cite{gay2} of recoil
polarization in elastic scattering of polarized electrons on
unpolarized protons at JLab have been used to extract data on the
ratio $\mu_pG_{Ep}(Q^2)/G_{Mp}(Q^2)$ for $0.49 GeV^2\leq Q^2\leq
5.54 GeV^2$, which disagree with Rosenbluth extractions from
cross-section measurements. Taking into account the dominance of
$G_{Mp}(t)$ in the unpolarized cross-section, we believe the
behaviour of $G_{Ep}(t)$ is responsible for this discrepancy. As a
result there are two sets of experimental data on nucleon EM FF's,
differing from each other by the different spacelike behaviour of
$G_{Ep}(t)$. Further we will predict from them strange nucleon
FF's behaviours by the specific eight-resonance unitary and
analytic model and look for distinctive features.

   The main idea \cite{jaf} of the prediction of strange nucleon FF's behaviours
is based on two assumptions:

\begin{itemize}
\item
on the $\omega-\phi$ mixing to be valid also for coupling
constants between EM (quark) current and vector-mesons

\begin{eqnarray}\nonumber
\frac{1}{f_\omega}&=& \frac{1}{f_{\omega_0}} \cos{\epsilon} -
\frac{1}{f_{\phi_0}} \sin{\epsilon}\\
\frac{1}{f_\phi}&=&\frac{1}{f_{\omega_0}} \sin{\epsilon} +
\frac{1}{f_{\phi_0}} \cos{\epsilon}
\end{eqnarray}

\item
on the assumption that the quark current of some flavour couples
with universal strength $\kappa$ exclusively to the vector-meson
wave function component of the same flavour
\begin{eqnarray}
\langle0|\bar q_r\gamma q_r|(\bar q_t q_t)_V\rangle = \kappa m^2_V
\delta_{rt}\varepsilon_\mu,
\end{eqnarray}
\end{itemize}
which result in the relations

\begin{eqnarray}
\nonumber ({f^{(i)}_{\omega
NN}}/{f^{s}_{\omega}})&=&-\sqrt{6}\frac{\sin{\epsilon}}
{\sin(\epsilon+\theta_{0})}({f^{(i)}_{\omega NN}}/{f^{e}_{\omega}})\\
({f^{(i)}_{\phi NN}}/{f^{s}_{\phi}})
&=&-\sqrt{6}\frac{\cos{\epsilon}}
{\cos(\epsilon+\theta_{0})}({f^{(i)}_{\phi NN}}/{f^{e}_{\phi}})
\label{CC}\\
\nonumber (i&=&1,2)
\end{eqnarray}
where $f^{s}_{\omega},f^{s}_{\phi}$ are strange-current
$\leftrightarrow V=\omega,\phi$ coupling constants and
$\epsilon=3.7^{0}$ is a deviation from the ideally mixing angle
$\theta_{0}=35.3^{0}$.

   So, if one knows from the fit of nucleon FF data free parameters
$(f^{(i)}_{\omega NN}/f^{e}_{\omega})$, $(f^{(i)}_{\phi
NN}/f^{e}_{\phi})$ (i=1,2) of the suitable eight-resonance model
for isoscalar parts of the Dirac and Pauli FF's
\begin{eqnarray}
\nonumber && F^{I=0}_{1}[V(t)]
 =(\frac{1-V^{2}}{1-V^{2}_{N}})^{4}\{\frac{1}{2}
 L(V_{\omega''})L(V_{\omega'})+\\
\nonumber && [L(V_{\omega''})L(V_{\omega})
 \frac{(C_{\omega''}-C_{\omega})}
 {(C_{\omega''}-C_{\omega'})}-
 L(V_{\omega'})L(V_{\omega})
 \frac{(C_{\omega'}-C_{\omega})}
 {(C_{\omega''}-C_{\omega'})}-\\
\nonumber && L(V_{\omega''})L(V_{\omega'})]
 (f^{(1)}_{\omega NN}/f^{e}_{\omega})+\\
\nonumber && [L(V_{\omega''})L(V_{\phi})
 \frac{(C_{\omega''}-C_{\phi})}
 {(C_{\omega''}-C_{\omega'})}-
 L(V_{\omega'})L(V_{\phi})
 \frac{(C_{\omega'}-C_{\phi})}
 {(C_{\omega''}-C_{\omega'})}-\\
&& L(V_{\omega''})L(V_{\omega'})]
 (f^{(1)}_{\phi NN}/f^{e}_{\phi})\}
\label{F01}
\end{eqnarray}

\begin{eqnarray}
\nonumber &&
F^{I=0}_{2}[V(t)]=(\frac{1-V^2}{1-V^2_N})^6\{L(V_{\omega''})
 L(V_{\omega'})L(V_{\omega})\\
\nonumber && [1-{\frac{C_{\omega}}{(C_{\omega''}-C_{\omega'})}}
 (\frac{(C_{\omega''}-C_{\omega})}{C_{\omega'}}-
 \frac{(C_{\omega'}-C_{\omega})}{C_{\omega''}})]\\
\nonumber && (f^{(2)}_{\omega NN}/f^{e}_{\omega})+
 L(V_{\omega''})L(V_{\omega'})L(V_{\phi})\\
\nonumber && [1-{\frac{C_{\phi}}{(C_{\omega''}-C_{\omega'})}}
 (\frac{(C_{\omega''}-C_{\phi})}{C_{\omega'}}-
 \frac{(C_{\omega'}-C_{\phi})}{C_{\omega''}})]\\
&& (f^{(2)}_{\phi NN}/f^{e}_{\phi})\} \label{F02}
\end{eqnarray}
with
\begin{eqnarray*}
\nonumber &&
L(V_r)=\frac{(V_N-V_r)(V_N-V^{\ast}_r)(V_N-1/V_r)(V_N-1/V^{\ast}_r)}
 {(V-V_r)(V-V^{\ast}_r)(V-1/V_r)(V-1/V^{\ast}_r)},\\
&& (r=\omega,\phi,\omega',\omega'')\\
\nonumber &&
C_r=\frac{(V_N-V_r)(V_N-V^{\ast}_r)(V_N-1/V_r)(V_N-1/V^{\ast}_r)}
 {-(V_r-1/V_r)(V^{\ast}_r-1/V^{\ast}_r)},\\
&& (r=\omega,\phi,\omega',\omega'')
\end{eqnarray*}
\begin{equation}
 V(t)=i\frac
 {\sqrt{[\frac{t_{N\bar N}-t^{I=0}_0}{t^{I=0}_0}]^{1/2}+
 [\frac{t-t^{I=0}_0}{t^{I=0}_0}]^{1/2}}-
 \sqrt{[\frac{t_{N\bar N}-t^{I=0}_0}{t^{I=0}_0}]^{1/2}-
 [\frac{t-t^{I=0}_0}{t^{I=0}_0}]^{1/2}}}
 {\sqrt{[\frac{t_{N\bar N}-t^{I=0}_0}{t^{I=0}_0}]^{1/2}+
 [\frac{t-t^{I=0}_0}{t^{I=0}_0}]^{1/2}}+
 \sqrt{[\frac{t_{N\bar N}-t^{I=0}_0}{t^{I=0}_0}]^{1/2}-
 [\frac{t-t^{I=0}_0}{t^{I=0}_0}]^{1/2}}}
\label{ITR}
\end{equation}

\begin{equation}
 V_N=V(t)_{|t=0}; V_r=V(t)_{|t=(m_r-i\Gamma_r/2)^2};
 (r=\omega,\phi,\omega',\omega''),
\label{DEF}
\end{equation}
and $t_{N\bar N}=4m_N^2$, then the unknown free parameters
$({f^{(i)}_{\omega NN}}/{f^{s}_{\omega}}), ({f^{(i)}_{\phi
NN}}/{f^{s}_{\phi}})$ of the strange nucleon FF's model
\begin{eqnarray}
\nonumber
&& F^{s}_{1}[V(t)]=(\frac{1-V^{2}}{1-V^{2}_{N}})^{4}\\
\nonumber && \{[L(V_{\omega''})L(V_{\omega})
 \frac{(C_{\omega''}-C_{\omega})}
 {(C_{\omega''}-C_{\omega'})}-
 L(V_{\omega'})L(V_{\omega})
 \frac{(C_{\omega'}-C_{\omega})}
 {(C_{\omega''}-C_{\omega'})}-\\
\nonumber && L(V_{\omega''})L(V_{\omega'})]
 (f^{(1)}_{\omega NN}/f^{s}_{\omega})+\\
\nonumber && [L(V_{\omega''})L(V_{\phi})
 \frac{(C_{\omega''}-C_{\phi})}
 {(C_{\omega''}-C_{\omega'})}-
 L(V_{\omega'})L(V_{\phi})
 \frac{(C_{\omega'}-C_{\phi})}
 {(C_{\omega''}-C_{\omega'})}-\\
&& L(V_{\omega''})L(V_{\omega'})]
 (f^{(1)}_{\phi NN}/f^{s}_{\phi})\}
\label{FS1}
\end{eqnarray}

\begin{eqnarray}
\nonumber &&
F^{s}_{2}[V(t)]=(\frac{1-V^2}{1-V^2_N})^6\{L(V_{\omega''})
 L(V_{\omega'})L(V_{\omega})\\
\nonumber && [1-{\frac{C_{\omega}}{(C_{\omega''}-C_{\omega'})}}
 (\frac{(C_{\omega''}-C_{\omega})}{C_{\omega'}}-
 \frac{(C_{\omega'}-C_{\omega})}{C_{\omega''}})]\\
\nonumber && (f^{(2)}_{\omega NN}/f^{s}_{\omega})+
 L(V_{\omega''})L(V_{\omega'})L(V_{\phi})\\
\nonumber && [1-{\frac{C_{\phi}}{(C_{\omega''}-C_{\omega'})}}
 (\frac{(C_{\omega''}-C_{\phi})}{C_{\omega'}}-
 \frac{(C_{\omega'}-C_{\phi})}{C_{\omega''}})]\\
&& (f^{(2)}_{\phi NN}/f^{s}_{\phi})\} \label{FS2}
\end{eqnarray}
of the same analytic structure, but with different normalization
of the Dirac FF, are calculated by the relations (\ref{CC}).

    The expressions (\ref{F01}) and (\ref{F02})
for $F_1^{I=0}, F_2^{I=0}$, together with similar expressions
\cite{dub} for $F_1^{I=1}, F_2^{I=1}$ to be saturated by $\rho,
\rho', \rho'', \rho'''$ isovector vector-mesons, have been used
\begin{itemize}
\item
first to describe Rosenbluth $G_{Ep}(t)$ data in $t<0$ region
together with all other existing nucleon EM FF data with the
result $\chi^2/(ndf)=1.76$
\item
then instead of the Rosenbluth $G_{Ep}(t)$ data in $t<0$ region
the JLab proton polarization data on
$\mu_pG_{Ep}(Q^2)/G_{Mp}(Q^2)$ for $-5.54 GeV^2\leq t \leq -0.49
GeV^2$ together with all other existing nucleon EM FF data were
analyzed with the result $\chi^2/(ndf)=1.34$.
\end{itemize}

\begin{figure}[htb]
\centerline{\includegraphics[width=0.45\textwidth]{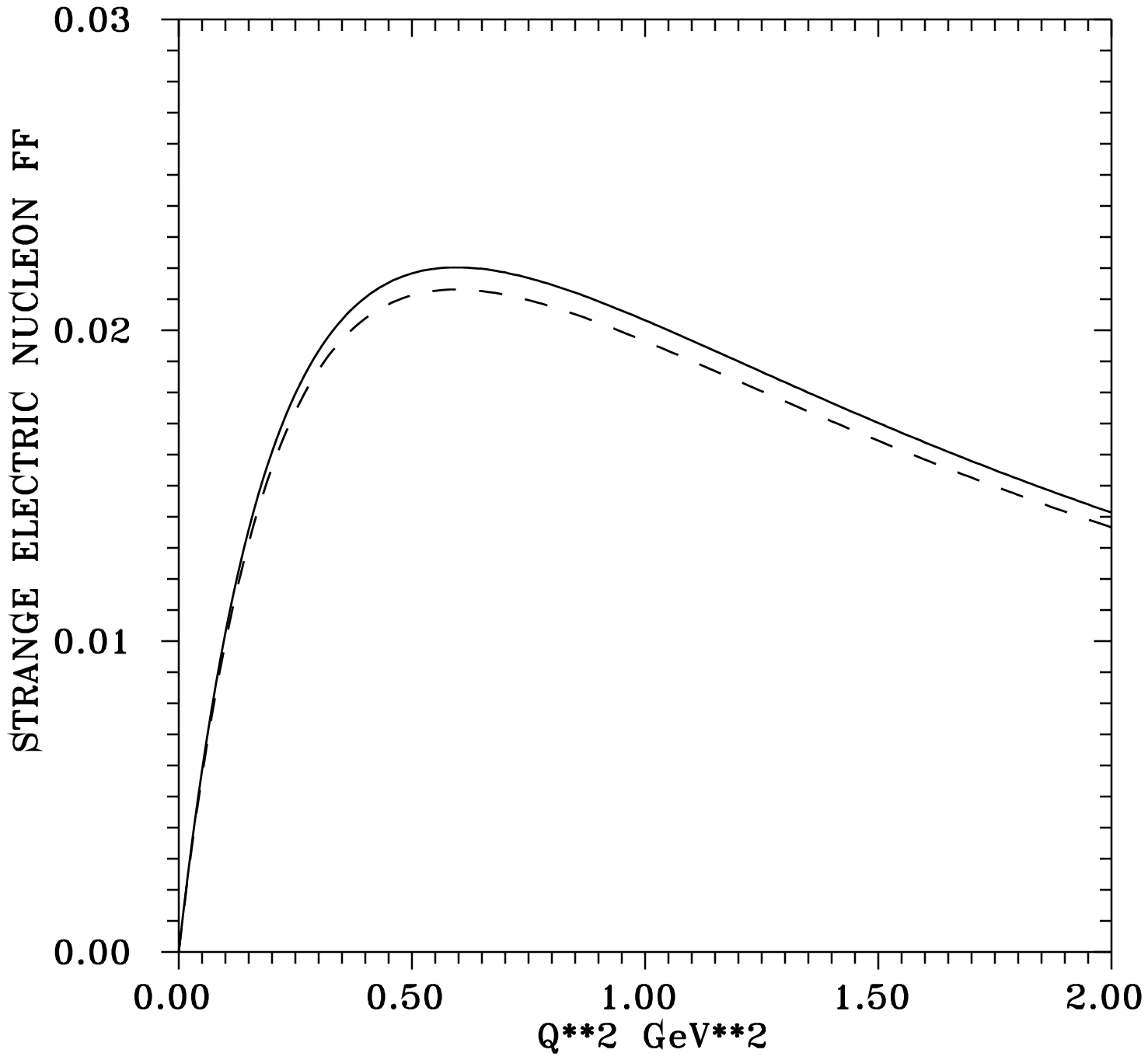}
\qquad
\includegraphics[width=0.45\textwidth]{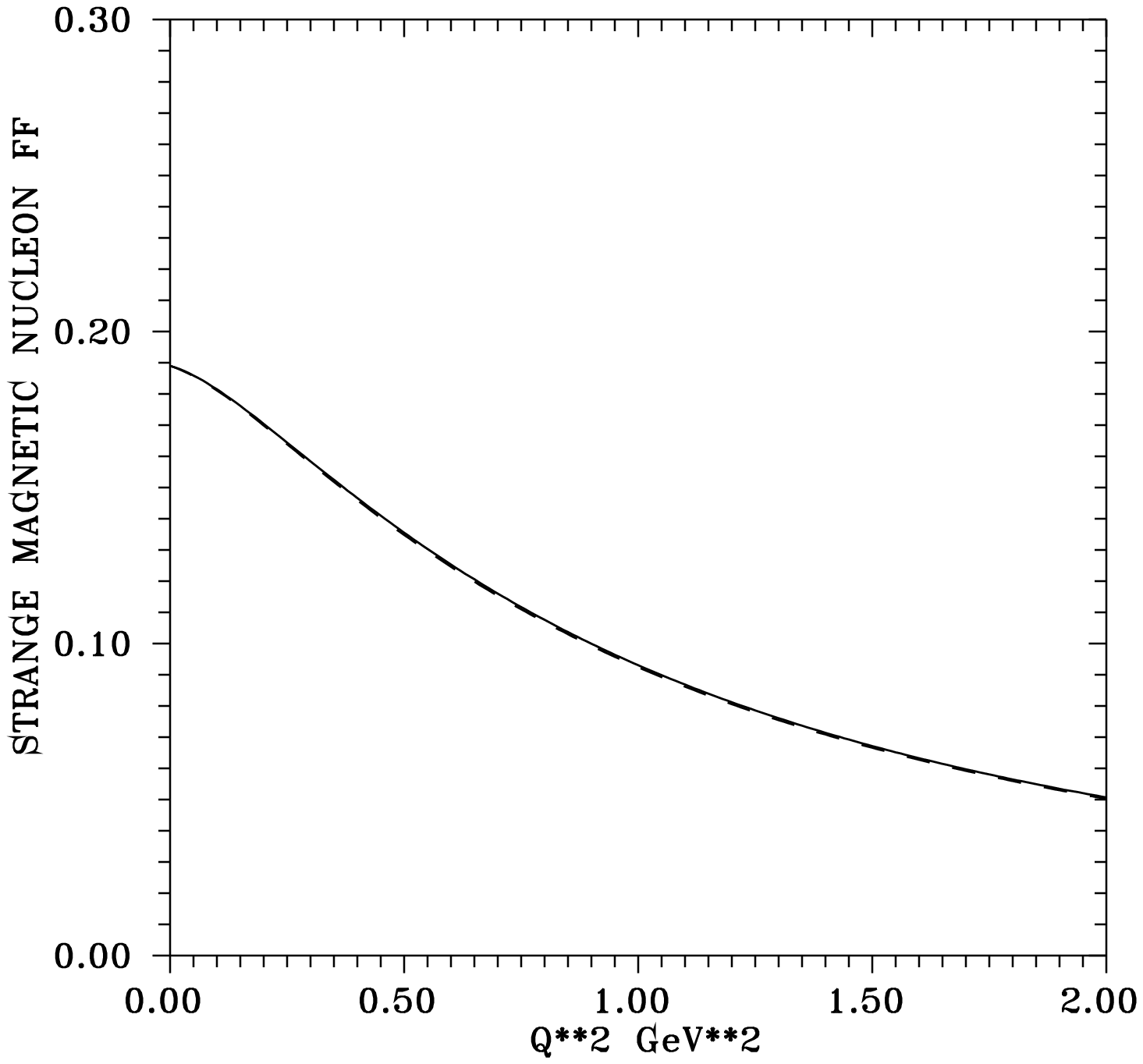}}
\caption{The predicted behaviors of strange nucleon electric and
magnetic form factors.}
\end{figure}

\begin{figure}[htb]
\centerline{\includegraphics[width=0.70\textwidth]{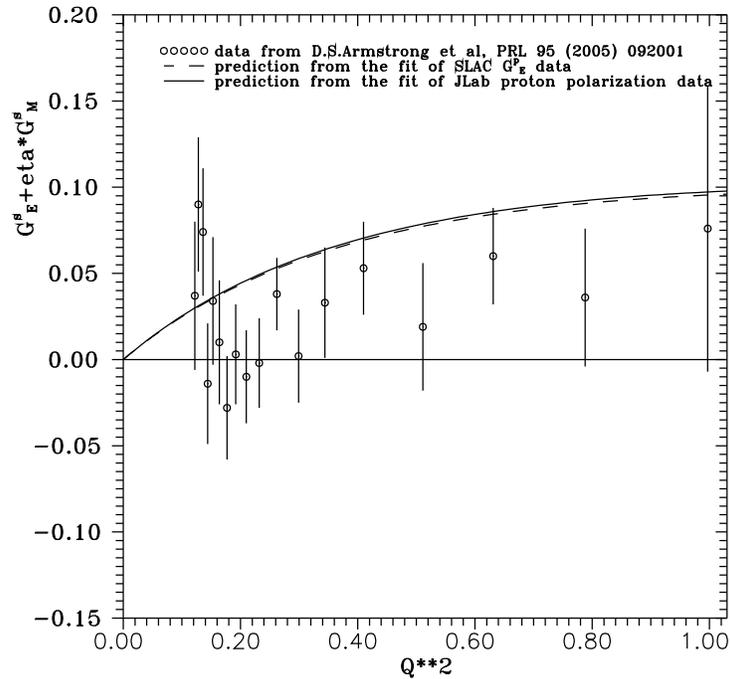}}
\caption{Predicted combination $G_E^s(Q^2)+\eta(Q^2)G_M^s(Q^2)$
from Rosenbluth and JLab data by 8-resonance U\&A model and its
comparison with the $G0$ collaboration data.}
\end{figure}

    Finally, the strange coupling constant ratios
$({f^{(i)}_{\omega NN}}/{f^{s}_{\omega}}), ({f^{(i)}_{\phi
NN}}/{f^{s}_{\phi}})$ according to the prescribed procedure given
by (\ref{CC}) were calculated and the behaviour of $G_E^s(Q^2),
G_M^s(Q^2)$ by means of (\ref{FS1}) and (\ref{FS2}), as presented
in Fig.1, are predicted. As one can see from Fig.1b, a reasonable
positive value of the strangeness nucleon magnetic moment is found
to be $\mu_s=+0.19 [\mu_N]$.

   A reasonable description of the recent data \cite{arm} on the
combination $G_E^s(Q^2)+\eta(Q^2)G_M^s(Q^2)$ for $0.12
GeV^2<Q^2<1.0GeV^2$ is achieved (see Fig.2) as well. Similar
results were obtained also by Bijker \cite{bij1} recently,
exploiting very simple parametrization \cite{bij2} of the nucleon
EM FF in the spacelike region.

   As one can see from Figs.1 and 2, the predicted behaviour of
the strange nucleon FF's by the special eight-resonance unitary
and analytic model doesn't feel too much the difference in
contradicting behaviours of $G_{Ep}(t)$ in the spacelike region.

\vspace{1cm}

   The work was in part supported by Slovak Grant Agency for Sciences, Gr. No
2/4099/26.

\end{document}